\documentclass[aps,prl,twocolumn,showpacs,superscriptaddress]{revtex4-1}

\usepackage[pdftex]{graphicx} 
\usepackage{natbib}
\usepackage{booktabs}

\newcommand{\one}{Fig.~\ref{f1}}
\newcommand{\two}{Fig.~\ref{f2}}
\newcommand{\three}{Fig.~\ref{f3}}
\newcommand{\four}{Fig.~\ref{f4}}
\newcommand{\sto}{SrTiO$_{3}$ }

\newcommand{\tg}{$t_{2g}$ }

\begin{document}

\title{Control of a two-dimensional electron gas on SrTiO$_3$(111) by atomic oxygen}

\author{S. McKeown Walker}
\affiliation{D\'epartement de Physique de la Mati\`ere Condens\'ee, Universit\'ee de Gen\`eve, 24 Quai Ernest-Ansermet, 1211 Gen\`eve 4, Switzerland}
\author{A. de la Torre}
\affiliation{D\'epartement de Physique de la Mati\`ere Condens\'ee, Universit\'ee de Gen\`eve, 24 Quai Ernest-Ansermet, 1211 Gen\`eve 4, Switzerland}
\author{F.Y. Bruno}
\affiliation{D\'epartement de Physique de la Mati\`ere Condens\'ee, Universit\'ee de Gen\`eve, 24 Quai Ernest-Ansermet, 1211 Gen\`eve 4, Switzerland}
\author{A. Tamai}
\affiliation{D\'epartement de Physique de la Mati\`ere Condens\'ee, Universit\'ee de Gen\`eve, 24 Quai Ernest-Ansermet, 1211 Gen\`eve 4, Switzerland}
\author{T. K. Kim}
\affiliation{Diamond Light Source, Harwell Campus, Didcot, OX11 0DE, United Kingdom}
\author{M. Hoesch}
\affiliation{Diamond Light Source, Harwell Campus, Didcot, OX11 0DE, United Kingdom}
\author{M. Shi}
\affiliation{Swiss Light Source, Paul Scherrer Institut, CH-5232 Villigen PSI, Switzerland}
\author{M.S. Bahramy}
\affiliation{Quantum-Phase Electronics Center, Department of Applied Physics, The University of Tokyo, Tokyo 113-8656, Japan}
\affiliation{RIKEN Center for Emergent Matter Science (CEMS), Wako 351-0198, Japan}
\author{P.D.C. King}
\affiliation{SUPA, School of Physics and Astronomy, University of St Andrews, St Andrews, Fife KY16 9SS, United Kingdom}
\author{F. Baumberger}
\affiliation{D\'epartement de Physique de la Mati\`ere Condens\'ee, Universit\'ee de Gen\`eve, 24 Quai Ernest-Ansermet, 1211 Gen\`eve 4, Switzerland}
\affiliation{Swiss Light Source, Paul Scherrer Institut, CH-5232 Villigen PSI, Switzerland}
\affiliation{SUPA, School of Physics and Astronomy, University of St Andrews, St Andrews, Fife KY16 9SS, United Kingdom}

\date{\today}

\begin{abstract}
{We report on the formation of a two-dimensional electron gas (2DEG) at the bare surface of (111) oriented SrTiO$_{3}$. Angle resolved photoemission experiments reveal highly itinerant carriers with a 6-fold symmetric Fermi surface and strongly anisotropic effective masses. The electronic structure of the 2DEG is in good agreement with self-consistent tight-binding supercell calculations that incorporate a confinement potential due to surface band bending. We further demonstrate that alternate exposure of the surface to ultraviolet light and atomic oxygen allows tuning of the carrier density and the complete suppression of the 2DEG. }
\end{abstract}

\maketitle
The emergent field of oxide electronics relies on the creation and manipulation of interface electronic states. Following the seminal discovery of a high mobility two-dimensional electron gas (2DEG) at the interface between the (001) oriented perovskite band insulators \sto (STO) and LaAlO$_3$ (LAO)~\cite{Ohtomo2004}, much work was devoted to revealing its unique magneto-transport properties, including gate controlled metal-insulator transitions~\cite{Caviglia2008}, superconductivity~\cite{Reyren2007} and its possible coexistence with magnetism~\cite{Brinkman2007}. Notably, 2DEGs in STO(001) can be created by very different means, such as bombardment of STO single crystals by Ar$^{+}$ ions \cite{Reagor2005}, electrolyte gating \cite{Ueno2008a} or deposition of amorphous \cite{Chen2011} and non-perovskite oxides \cite{Chen2013a}. Yet, the resulting surface and interface 2DEGs all display similar electronic transport phenomena, suggesting a common underlying electronic structure defined by the properties of STO and the crystallographic orientation of the surface or interface. Recent angle resolved photoemission (ARPES) experiments from the bare STO(001) surface \cite{Meevasana2011a, Santander-Syro2011a, King2012a, Santander-Syro2012, King2014} indeed indicate a crucial role of the confinement direction in shaping key-properties of the 2DEG there such as orbital ordering and, related to this, an unconventional Rashba splitting, possibly driving the marked density dependence of magneto-transport phenomena in STO(001) 2DEGs \cite{Joshua2012,King2014,Zhong2013b, Khalsa2013a,Fete2012,Kimura2004,Nakamura2012}. 

\begin{figure}[!t]
\includegraphics[width=8.5cm]{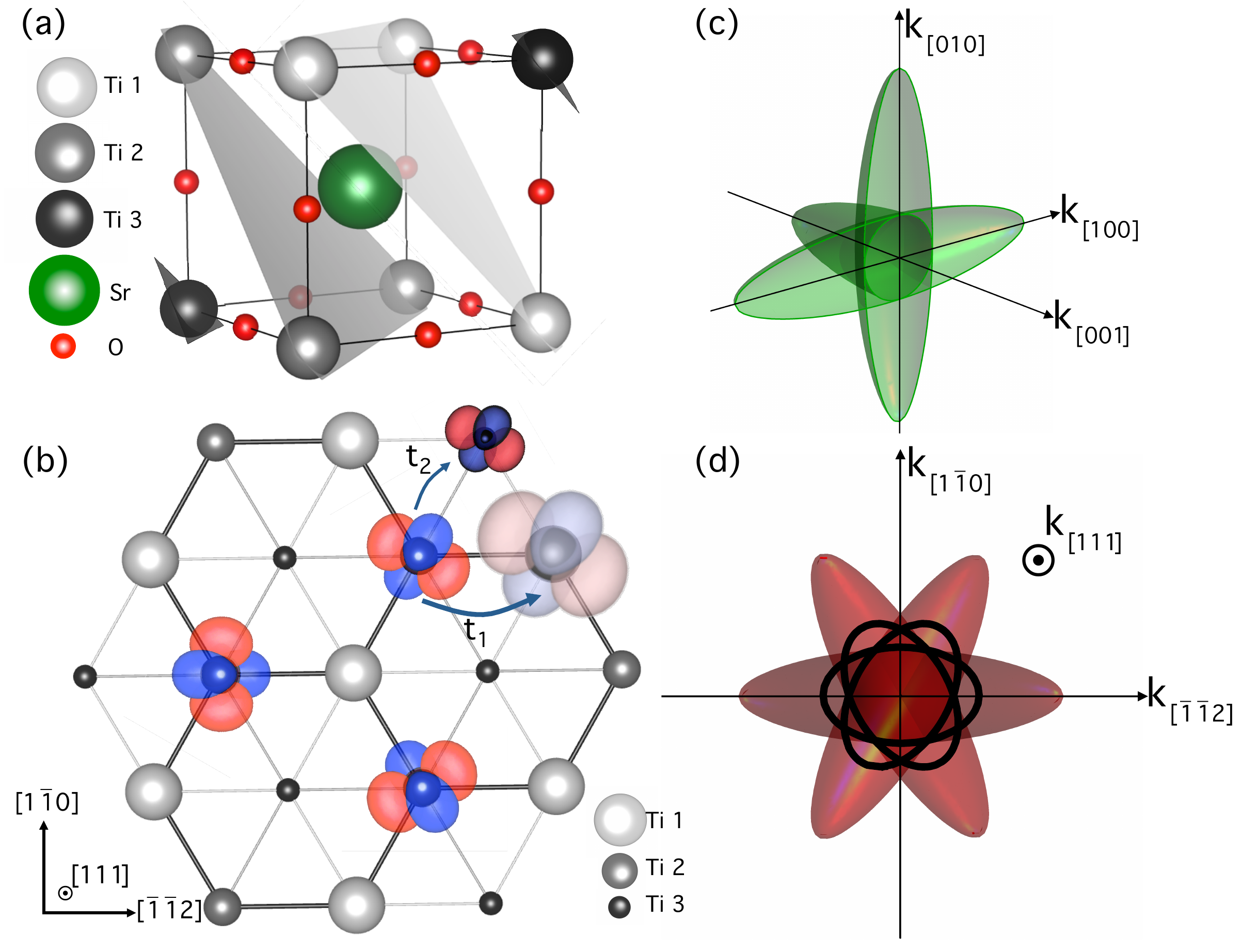} %
\caption{(color online) (a) Schematic of a cubic perovskite unit cell of \sto with inequivalent (111) planes indicated in grey. (b) Top view of three consecutive Ti$^{4+}$ (111) layers. Each of the three \tg orbitals are shown on the honeycomb lattice, formed by two consecutive layers (dark grey lines), to illustrate their rotational symmetry. The $xz$ orbital is shown in all three layers with large/small (t$_{1}$/t$_{2}$) nearest neighbour hoppings indicated. (c) Sketch of the bulk FS of STO cut by the (001) plane. (d) The same bulk FS viewed down the [111] axis. A cut in the (111) plane through the $\Gamma$ point is indicated by black lines to illustrate its different shape and size from the projection of the FS on the (111) surface plane.}
\label{f1}
\end{figure}

It is thus natural to investigate different surface terminations in the search for 2DEGs with novel and potentially useful properties. Recent theoretical work suggests that (111) oriented perovskites might display particularly intriguing phenomena. Along this direction the ABO$_3$ cubic perovskite structure can be considered as a stacking of ionic AO$_{3}$ and B-site planes with the B-site ions forming a triangular lattice (see \one). Two consecutive B-site planes will thus form a honeycomb lattice potentially suitable for realizing novel topological phases \cite{Xiao2011,Yang2011,Bareille2014b}. However, despite the successful creation of 2DEGs at the(111)-oriented interface of STO/LAO~\cite{Herranz2012} and at the bare surface of KTaO$_3$~\cite{Bareille2014b}, little is known to date about their origin and microscopic electronic structure.\\

Here we use angle resolved photoemission to show that the {\it in-situ} cleaved (111) surface of \sto supports a robust, quantum confined two-dimensional electron system. Our model calculation fully reproduces the observed electronic structure and describes how the 2DEG emerges from the quantum confinement of \tg electrons near the surface due to band bending. Moreover, we demonstrate how the 2DEG can be reversibly depleted and created by alternately exposing the surface to low doses of atomic oxygen and ultraviolet (UV) light, providing insight into the origin of the confining potential. 

Single crystals of commercially grown (Crystal GmbH), lightly electron doped SrTi$_{1-  x}$Nb$_{ x}$O$_{3}$(111) (\( x  = 0.002\)) were measured. The Nb doping results in a small residual bulk conductivity and a maximum doping of  $3 \times 10^{19} $ cm$^{-3}$,  which  helps to eliminate charging effects during ARPES but does not otherwise influence our results \cite{spinelli2010}. Samples were cleaved at a pressure lower than \(5 \times 10 ^{-10} \) mbar at the measurement temperature. ARPES measurements (T \(= 20\) K, \(hv = 40 - 140 \) eV) were performed at the SIS beamline of the Swiss Light Source and the I05 beamline of the Diamond Light Source with an angular resolution of $\approx0.2$\(^\circ \) and an energy resolution of 10 - 20 meV.

\begin{figure}[!t]
\includegraphics[width=8.5cm]{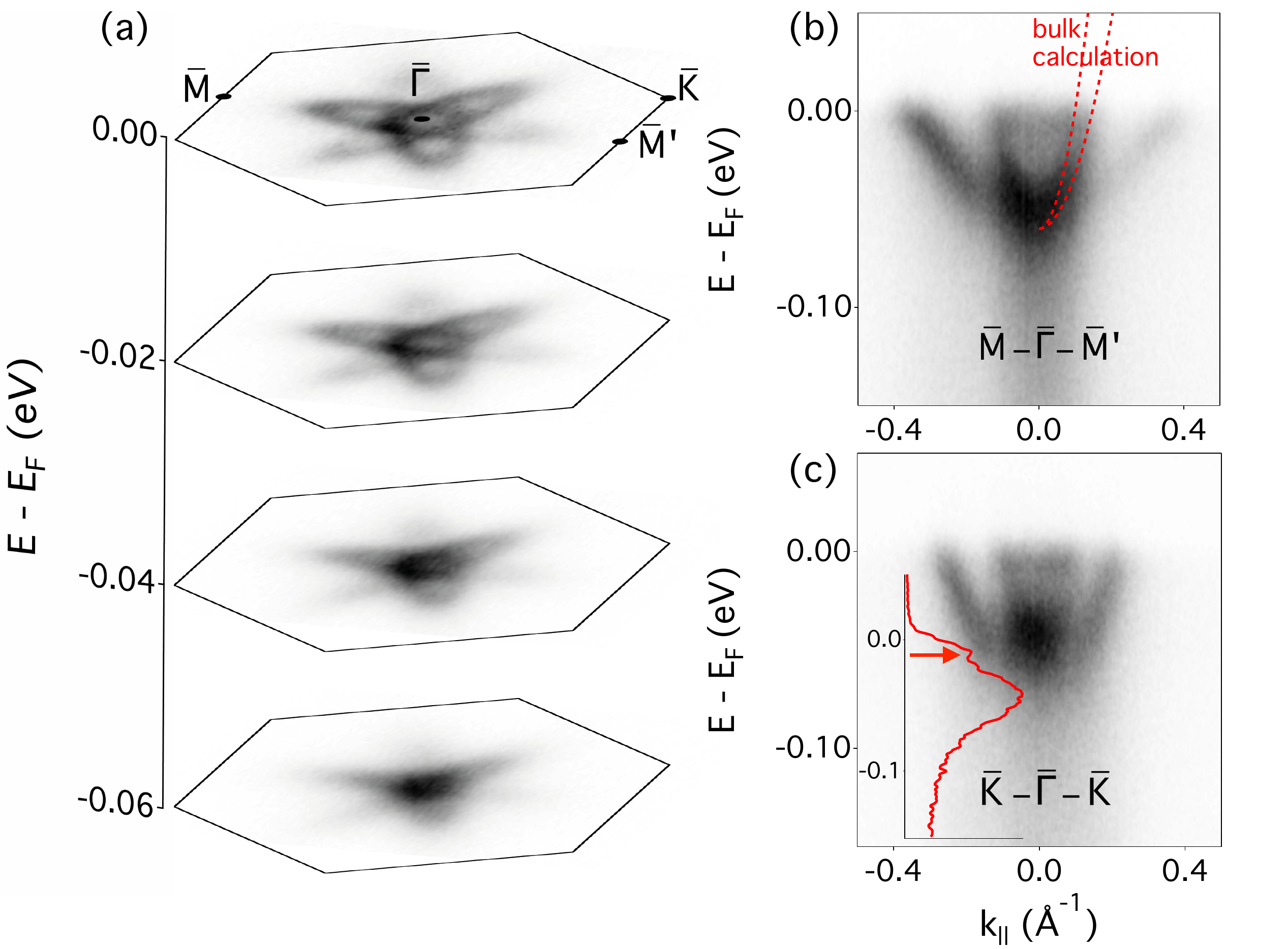}
\caption{(color online) (a) Constant energy surfaces of the STO(111) electronic structure measured at 108~eV with circularly polarized light. (b,c) Energy - momentum dispersion along two high-symmetry directions (108~eV, $s$-polarization). A calculation of the bulk dispersion in the same direction is indicated by red dashed lines in (b). Inset of (c) is an EDC at $\Gamma$, the red arrow indicates the peak due to the second subband.}
\label{f2}
\end{figure}

Figure 1 illustrates the different symmetry and atomic arrangement of STO(111) and STO(001) surfaces. While the $xy$ and $xz/yz$ orbitals have very different overlap along the [001] confinement direction of STO(001), all \tg orbitals are equivalent modulo a rotation of 120$^{\circ}$ when viewed along the surface normal of STO(111) [see \one(b)]. This immediately suggests that quantum confinement will not drive a significant orbital polarization in the STO(111) 2DEG, in stark contrast to STO(001)~\cite{Popovic2008,Salluzzo2009,Zhong2013b,Khalsa2013a,King2014}, which is indeed what we observe experimentally.

The ARPES Fermi surface [FS] shown in \two{}(a) consists of three equivalent elliptical sheets oriented along $\bar\Gamma\bar\mathrm{M}$, consistent with the 3-fold rotational symmetry of the surface \footnote{Since we observe that the matrix elements of photoemission process have a lower symmetry when linearly polarized light is used, the data presented in figure \two(a) is the sum of spectra taken with right- [C$^{+}$] and left-handed [C$^{-}$] circularly polarized light.}. 
The band structure along $\bar\Gamma\bar\mathrm{M}$ [\two(b)] shows a single heavy band, corresponding to the long axis of one of the FS ellipses, which is nearly degenerate at the band bottom with a more dispersive, doubly degenerate band arising from the two remaining FS sheets. Near the Fermi level, additional dispersive spectral weight is observed, indicating a second occupied subband, which is a natural consequence of quantum confinement and is not observed in the bulk. We also confirmed the 2D nature of the charge carriers directly through extensive photon energy dependent measurements (not shown) which reveal no signs of dispersion along $k_z$ within the resolution of the experiment. From the FS area we deduce a 2D carrier density of $1.5 \times 10^{14}$~cm$^{-2}$. Assuming a constant electron density over approximately 15 Ti layers (33~\AA) below the surface, this would correspond to to a 3D density at the surface of $4.5 \times 10^{20}$~cm$^{-3}$, more than an order of magnitude higher than the nominal bulk doping of our samples. Together, these observations conclusively demonstrate that the observed electronic structure arises from quantum confinement in a surface 2DEG.

Intriguingly, the FS-ellipses we observe are elongated with respect to those of our bulk electronic structure calculations in the (111) plane as indicated by thin red lines in \two(b). Fitting the measured dispersion over an extended energy range, we find band masses of $\bar m_{L} \approx 0.8$~m$_e$ and $\bar m_{H} \approx 8.7$~m$_e$ for the short and long axes of the FS-ellipses respectively, whereas the calculated bulk band masses along the same crystallographic directions are $\approx0.6$~m$_{e}$ and $\approx1.5$~m$_e$, respectively. This indicates a remarkable reconstruction of the electronic structure, in stark contrast to the case of the STO(001) 2DEG, where the band masses are similar to the bulk except for a moderate low-energy renormalization due to electron-phonon coupling~\cite{King2014}.

We first discuss these observations on a qualitative level. Since in a 2D system the electronic states have the same energy for all values of the momentum $k_z$ perpendicular to the surface, the Fermi surface of the 2DEG can be approximated by a projection of the 3D bulk band structure (at appropriate doping levels) onto the surface plane. In the case of STO(001), the surface projection is identical to a cut through the bulk FS as is evident from the sketch in \one(c). Consequently the band masses of the 2DEG closely resemble those of the bulk calculations. In contrast, a projection onto the (111) plane results in elliptical contours that are strongly elongated compared to a cut through the bulk FS as demonstrated in \one(d), consistent with our observation of enhanced effective masses. Additionally, individual subbands can have different confinement energies $\epsilon$ depending on their effective masses for motion perpendicular to the surface. Given that $\epsilon\propto m_{\!\perp}^{-\alpha}$ where $\alpha=1$ for a square- and $\alpha\approx\frac{1}{3}$ for a triangular quantum well respectively, it is immediately clear that orbitals with a large hopping amplitude along the confinement direction will sink deepest in the quantum well. This effect dominates the large lifting of the degeneracy of $\textit{xy}$ and $\textit{xz/yz}$ bands observed in the STO(001) 2DEG~\cite{Popovic2008,Salluzzo2009,Zhong2013b,Khalsa2013a,King2014}. In the STO(111) 2DEG, on the other hand, all \tg orbitals have the same effective mass along the surface normal of $m_{[111]}\approx 0.8$~m$_e$. Hence, the three \tg orbitals are nearly degenerate at the $\Gamma$-point and have a shallow bandwidth comparable to the $\textit{xz/yz}$ sheets in the (001) 2DEG.

\begin{figure}[!t]
\includegraphics[width=9cm]{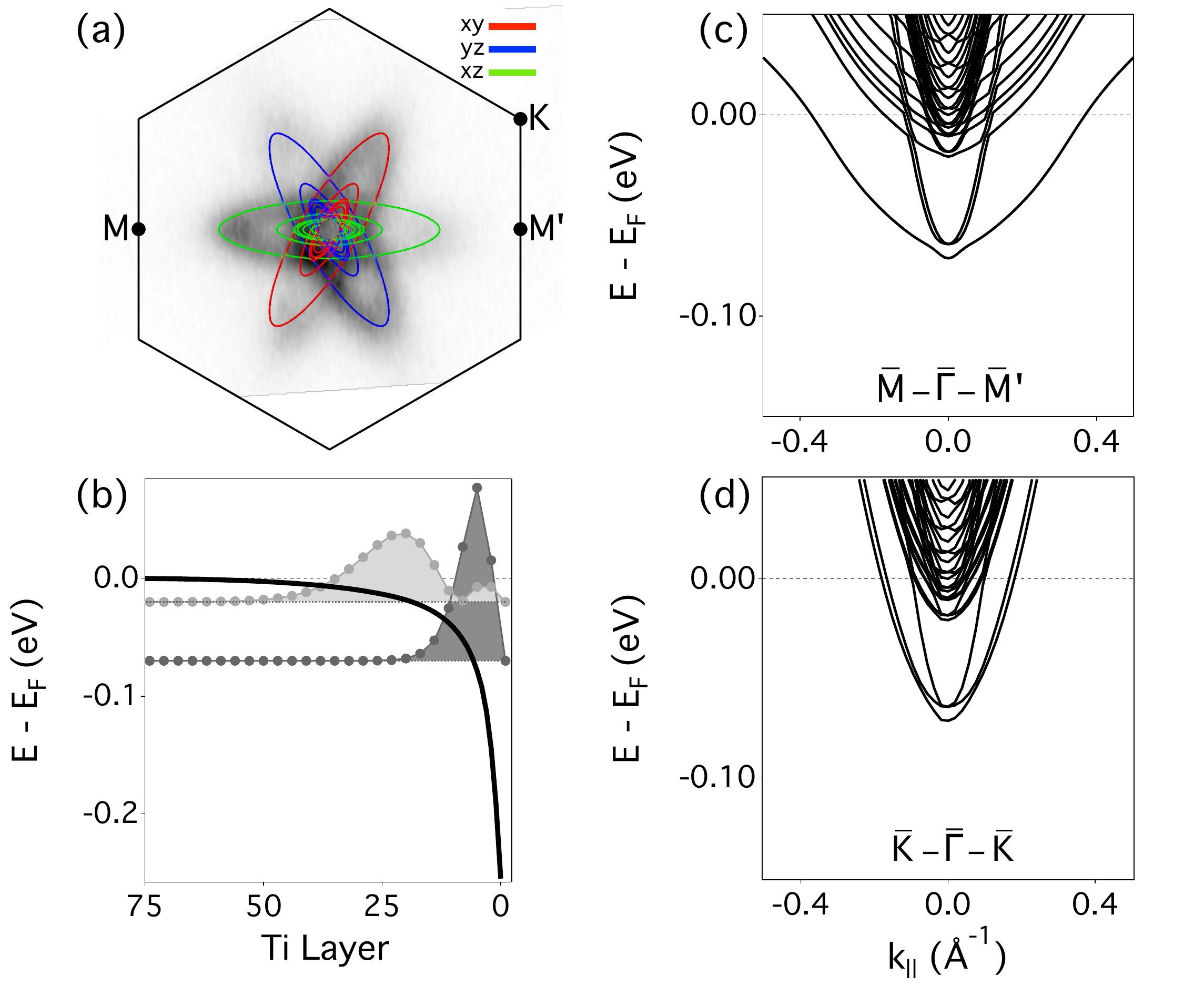}
\caption{(color online) Tight-binding supercell calculations of the electronic structure at the (111) surface of SrTiO$_3$. (a) The calculated FS and orbital character superposed on the ARPES data. (b) The self-consistent band bending potential. One Ti layer corresponds to 2.25~\AA. The square modulus of the wave functions of the lowest two subbands at $\Gamma$ are plotted at their corresponding confinement energies. (c,d) Calculated band dispersion along two high-symmetery directions, showing three confined 2DEG subbands and a 'ladder' of states above E$_{F}$ due to the finite size of the supercell.}
\label{f3}
\end{figure}

\begin{figure*}[!t]
\includegraphics[width=18cm]{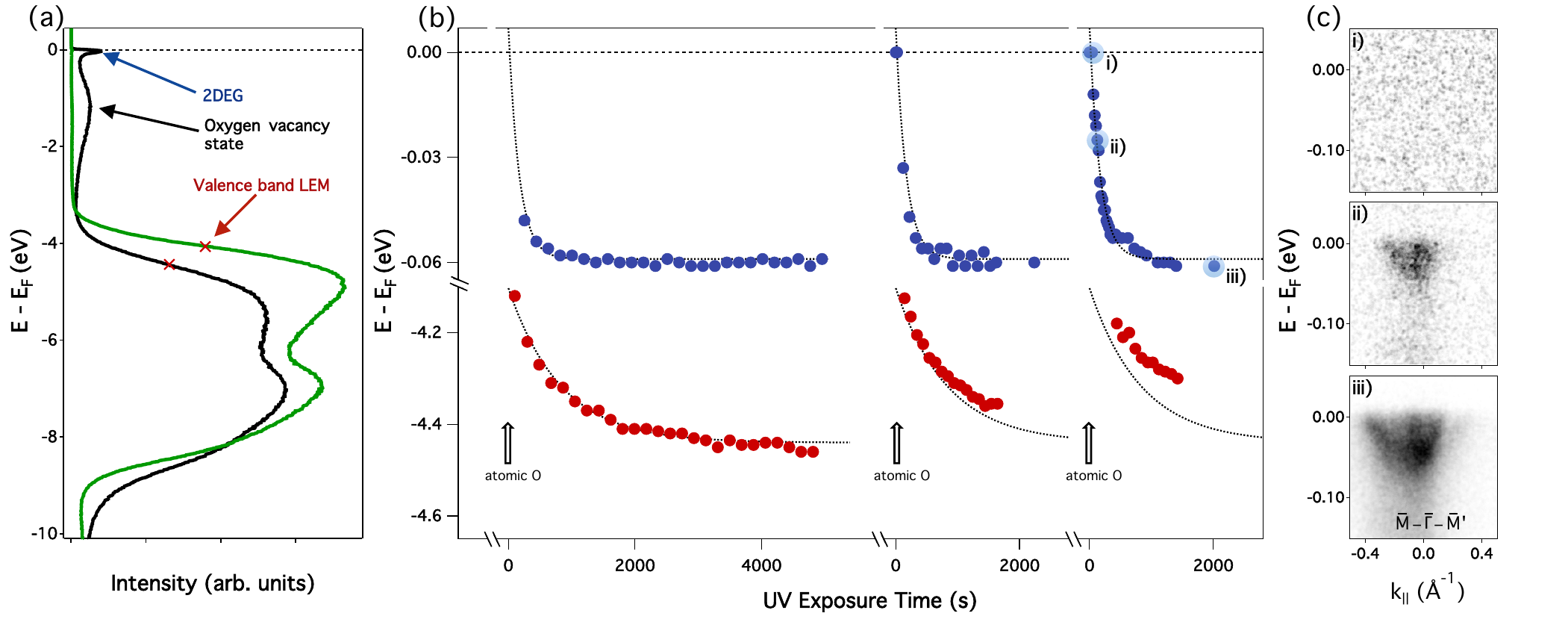}
\caption{(color online) Atomic oxygen treatment of the cleaved STO(111) surface. (a) Angle integrated photoemission spectra from a surface with a fully developed/suppressed 2DEG after UV/atomic oxygen (black/green) exposure. (b) Occupied bandwidth of the 2DEG (blue) and valence band leading-edge mid-point (red) as a function of time exposed to UV radiation after atomic oxygen exposure which is indicated by arrows and marked by a break in the time axis. The dashed black lines are guides to the eye. Zero bandwidth indicates complete suppression of the 2DEG. (c) Dispersion plots in the $\bar\textrm{M'} -\bar\Gamma-\bar\textrm{M}$ direction from three stages of the 2DEG development as indicated in panel (b).}
\label{f4}
\end{figure*}

This qualitative understanding is supported by tight-binding supercell calculations based on a non-relativistc {\it ab initio} bulk band structure.
To calculate the surface electronic structure, a supercell containing 120 Ti atoms stacking along [111] direction was constructed using maximally localized Wannier functions with additional on-site potential terms to account for band bending via an electrostatic potential variation. The tight-binding hamiltonian was solved self-consistently with Poisson's equation, incorporating an electric field dependent dielectric constant \cite{Copie2009,Bahramy2012,King2014}. The only adjustable parameter is the total magnitude of the band bending at the surface, which we choose to reproduce the experimentally observed binding energy of the lowest subband ($\approx60$ meV). For simplicity, the calculation assumes a bulk truncated (111) surface devoid of any reconstructions. This is supported by the experimentally observed symmetry of the FS and the absence of backfolded bands, which we confirmed for different photon energies and over multiple Brillouin zones.

The results of this calculation clearly reproduce the experimentally observed band dispersion and Fermi surface of the lowest subband as shown in \three. They further indicate a ladder of three higher subbands with progressively more bulk like character. The second subband is predicted just below the Fermi energy, again in good agreement with the experiment. The wave functions of the lowest subband at $\Gamma$ extend over 15 - 25 Ti layers (34 - 56~\AA), and thus nearly an order of magnitude more than the bulk penetration of the first \textit{xy} subband on STO(001)~\cite{King2014}. This can be attributed to the lighter effective mass perpendicular to the surface and the correspondingly shallower binding energy. The orbital character of the FS shown in \three(a) confirms the basic picture that each \tg orbital contributes an equivalent elliptical Fermi surface sheet.

For the remainder of this paper, we discuss the origin of the 2DEG and its relation to interface 2DEGs. Since STO is a band insulator, charge carriers and appropriate electrostatic boundary conditions, namely band bending, are needed to create a 2DEG. In the ideal polar catastrophe scenario for LAO/STO(001), both are provided by the intrinsic electric field in the LAO layer, which allows separation of electrons and holes, resulting in electron accumulation in STO and doped holes on the surface of LAO. While there is strong experimental evidence for this scenario in the case of LAO/STO(001)~\cite{Nakagawa2006,Reinle-Schmitt2012}, it clearly cannot apply to a bare STO surface. 

Experimentally, we observe that the 2DEG develops on the STO(111) surface only after exposure to intense ultraviolet radiation, just as for the 2DEG of STO(001)~\cite{Meevasana2011a,King2014}, indicating a common microscopic origin of the quantum confined charge carriers on both surfaces. For the case of STO(001), we proposed earlier that light-induced oxygen desorption might drive the formation of the 2DEG~\cite{Meevasana2011a}. At low temperature, vacancy diffusion will be strongly suppressed and oxygen desorption will be limited to the topmost layer. The system will then try to screen the positively charged vacancies with mobile carriers, which requires band bending such that the edge of the conduction band is dragged below the Fermi level in a narrow layer at the surface. 

To test this hypothesis we exposed the cleaved surface of STO at low temperature ($\approx 8$~K) alternately to UV light and atomic oxygen while monitoring the presence of the 2DEG and changes in the O~$2p$ valence bands. The angle integrated photoemission spectra of the surface after illumination with the synchrotron UV beam ($h\nu=108$~eV, $\approx10^{15}$~photons~s$^{-1}$mm$^{-2}$) for $\approx1$~h (black) and immediately after exposure to atomic oxygen (green) are shown in \four(a) \footnote{The samples were exposed for $\approx 30$~s to a thermal gas cracker operated at $\approx 1700$~$^{\circ}$C and a flow of 0.004~sccm ($1\times 10^{-7}$~mbar chamber pressure) resulting in a mixture of $\approx 9\%$ atomic and $\approx 91\%$ molecular oxygen}. The main features of the spectra are the sharp 2DEG peak just below the Fermi level, the O~$2p$ valence band with an onset near $-3$~eV and an in-gap state around $-1.3$~eV, which was attributed to oxygen vacancies in Refs.~\cite{Meevasana2010,Aiura2002,Lechermann2014,Lin2013}. Upon exposure to atomic oxygen the 2DEG peak and the in-gap defect state completely vanish, and concomitant with the depletion of the 2DEG, the VB leading edge mid-point [LEM] shifts towards the Fermi level by $\approx300$ meV. Assuming a constant band-gap, this implies a strongly reduced or vanishing band bending following exposure to atomic oxygen. We note that the exact magnitude of the surface band bending is difficult to extract from such data since the photoemission probing depth is comparable to the width of surface band bending. The VB peak will thus contain contributions from unit cells with locally different band energies.

\four(b) shows the systematic time evolution of the valence band leading edge and the occupied 2DEG bandwidth as the surface is repeatedly exposed to atomic oxygen and UV radiation. The data clearly show that the onset of band bending coincides with near-surface charge accumulation and that both, the bandwidth of the 2DEG, and the depth of the confining potential saturate at long time \footnote{To account for a slight angular misalignment in the first run shown in \four(b), we corrected the VB leading edge mid-point and 2DEG bandwidth by 23~meV and 40~meV, respectively.}. In parallel to this, the in-gap state loses all spectral weight when the surface is treated with atomic oxygen, consistent with the filling of all oxygen vacancies, and recovers as the surface is irradiated again. 
In \four(c) we show three dispersion plots obtained after the third consecutive exposure to atomic oxygen. Immediately after oxygen exposure the surface is insulating and no spectral weight is observed at the Fermi level. After $\approx100$~s UV exposure a 2DEG with an occupied bandwidth around 30~meV develops. Following longer exposure to UV light, the 2DEG fully recovers the bandwidth and Fermi wave vector of samples that were not treated with atomic oxygen. This indicates that the electronic structure of the 2DEG is largely defined by the electrostatic boundary conditions controlled by the density of charged oxygen vacancies on the surface and is remarkably insensitive to the detailed atomic structure of the surface, which likely changes after exposure to atomic oxygen. We attribute this to the spatial extent of the wave function, which has little weight in the topmost plane and peaks around 6 Ti layers below the surface [see \three(b)].
 
Together, our observations conclusively demonstrate that the STO(111) surface supports a 2DEG with 6-fold symmetry, which is quantum confined by a band bending potential induced by surface oxygen vacancies.
We further demonstrated control of the 2DEG bandwidth by alternate exposure of the surface to UV light and atomic oxygen providing a broadly applicable route to controlling the carrier density and thus the macroscopic properties of transition metal oxide surface 2DEGs.

\begin{acknowledgments}
We gratefully acknowledge discussions with D. Van der Marel and J.-M. Triscone.
This work was supported by Swiss National Science Foundation (200021-146995). PDCK was supported by the UK-EPSRC and MSB by by Grant-in-Aid for Scientific Research (S) (No. 24224009) from the Ministry of Education, Culture, Sports, Science and Technology (MEXT) of Japan.
\end{acknowledgments}

\bibliography{Bib_v6.bib}

\end{document}